# The CERN n_TOF NEAR station for astrophysics- and application-related neutron activation measurements.

N. Patronis[1,2], A. Mengoni[2,3], N. Colonna[4a], M. Cecchetto[2], C. Domingo-Pardo[5], O. Aberle[2], J. Lerendegui-Marcos[5], G. Gervino[6], M.E. Stamati[1], S. Goula[1], A.P. Bernardes[2], M. Mastromarco[4,41], A. Manna[7], R. Vlastou[8], C. Massimi[7,40], M. Calviani[2], V. Alcayne[9], S. Altieri[10], S. Amaducci[11], J. Andrzejewski[12], V. Babiano-Suarez[5], M. Bacak[2], J. Balibrea[5], C. Beltrami[10], S. Bennett[13], E. Berthoumieux[14], M. Boromiza[15], D. Bosnar[16], M. Caamaño[17], F. Calviño[18], D. Cano-Ott[9], A. Casanovas[18], F. Cerutti[2], G. Cescutti[19,20], S. Chasapoglou[8], E. Chiaveri[2], P. Colombetti[21], P. Console Camprini[7], G. Cortés[18], M. A. Cortés-Giraldo[22], L. Cosentino[11], S. Cristallo[23,24], S. Dellmann[25], M. Di Castro[2], S. Di Maria[26], M. Diakaki[8], M. Dietz[27], R. Dressler[28], E. Dupont[14], I. Durán[17], Z. Eleme[1], S. Fargier[2], B. Fernández[25], B. Fernández-Domínguez[17], P. Finocchiaro[11], S. Fiore[29], V. Furman[30], F. García-Infantes[31], A. Gawlik-Ramięga[12], S. Gilardoni[2], E. González-Romero[9], C. Guerrero[22], F. Gunsing[14], C. Gustavino[31], J. Heyse[33], W. Hillman[12], D. G. Jenkins[34], E. Jericha[35], A. Junghans[36], Y. Kadi[2], K. Kaperoni[8], G. Kaur[14], A. Kimura[37], I. Knapová[38], M. Kokkoris[8], Y. Kopatch[30], M. Krtička[38], N. Kyritsis[8], I. Ladarescu[2], C. Lederer-Woods[39], G. Lerner[2], T. Martínez[9], A. Masi[2], P. Mastinu[41], E. A. Maugeri[28], A. Mazzone[43], E. Mendoza[9], P. M. Milazzo[19], R. Mucciola[23], F. Murtas[44], E. Musacchio-Gonzalez[41], A. Musumarra[45], A. Negret[15], A. Pérez de Rada[9], P. Pérez-Maroto[22], J. A. Pavón-Rodríguez[22], M. G. Pellegriti[7], J. Perkowski[12], C. Petrone[15], E. Pirovano[27], J. Plaza[2], S. Pomp[46], I. Porras[31], J. Praena[31], J. M. Quesada[22], R. Reifarth[25], D. Rochman[28], Y. Romanets[26], C. Rubbia[2], A. Sánchez[9], M. Sabaté-Gilarte[2], P. Schillebeeckx[33], D. Schumann[28], A. Sekhar[13], A. G. Smith[13], N. V. Sosnin[39], A. Sturniolo[6], G. Tagliente[4], D. Tarrío[46], P. Torres-Sánchez[31], S. Urlass[36], E. Vagena[1], S. Valenta[38], V. Variale[4], P. Vaz[26], G. Vecchio[11], D. Vescovi[25], V. Vlachoudis[2], T. Wallner[36], P. J. Woods[39], T. Wright[13], R. Zarrella[40], and P. Žugec[16] The n_TOF Collaboration

[1]University of Ioannina, Greece; [2]CERN, Switzerland; [3]ENEA, Bologna, Italy; [4]INFN, Sez. di Bari, Italy; [5]Instituto de Física Corpuscular, CSIC - Universidad de Valencia, Spain; [6]INFN, Sez. di Torino, Italy; [7]INFN, Sez. di Bologna, Italy; [8]National Technical University of Athens, Greece; [9]Centro de Investigaciones Energéticas Medioambientales y Tecnológicas (CIEMAT), Spain; [10]INFN and Università di Pavia, Pavia, Italy; [11]INFN - Laboratori Nazionali del Sud, Catania, Italy; [12]University of Lodz, Poland; [13]University of Manchester, United Kingdom; [14]CEA Irfu, Université Paris-Saclay, F-91191 Gif-sur-Yvette, France; [15]Horia Hulubei National Institute of Physics and Nuclear Engineering, Romania; [16]Department of Physics, Faculty of Science, University of Zagreb, Zagreb, Croatia; [17]University of Santiago de Compostela, Spain; [18]Universitat Politècnica de Catalunya, Spain; [19]INFN, Sez. di Trieste, Italy; [20]INAF - Osservatorio Astronomico di Trieste, Italy; [21]INFN, Sez. di Torino, Italy; [22]Universidad de Sevilla, Spain; [23]INFN, Sez. di Perugia, Italy; [24]INAF - Osservatorio Astronomico di Teramo, Italy; [25]Goethe University Frankfurt, Germany; [26]Instituto Superior Técnico, Lisbon, Portugal; [27]Physikalisch-TechnischeBundesanstalt (PTB), Bundesallee 100, 38116 Braunschweig, Germany; [28]Paul Scherrer Institut (PSI), Villigen, Switzerland; [29]ENEA, Frascati, Italy; [30]Joint Institute for Nuclear Research (JINR), Dubna, Russia; [31]University of Granada, Spain; [32]INFN, Sez. di Roma, Italy; [33]European Commission, Joint Research Centre (JRC), Geel, Belgium; [34]University of York, United Kingdom; [35]TU Wien, Atominstitut, Stadionallee 2, 1020 Wien, Austria; [36]Helmholtz-Zentrum Dresden-Rossendorf, Germany; [37]Japan Atomic Energy Agency (JAEA), Tokai-Mura, Japan; [38]Charles University, Prague, Czech Republic; [39]School of Physics and Astronomy, University of Edinburgh, United Kingdom; [40]Dip. Fisica e Astronomia, Università di Bologna, Italy; [41]INFN - Laboratori Nazionali Legnaro, Italy; [42]Dip. Interateneo di Fisica, Università di Bari, Italy; [43]CNR - Bari, Italy; [44]INFN - Laboratori Nazionali di Frascati, Italy; [45]Dip- di Fisica, Università di Catania, Italy; [46]Uppsala University, Sweden

1
2

**Abstract.** A new experimental area, the NEAR station, has recently been built at the CERN n_TOF facility, at a short distance from the spallation target (1.5 m). The new area, characterized by a neutron beam of very high flux, has been designed with the purpose of performing activation measurements of interest for astrophysics and various applications. The beam is transported from the spallation target to the NEAR station through a hole in the shielding wall of the target, inside which a collimator is inserted. The new area is complemented with a γ-ray spectroscopy laboratory, the GEAR station, equipped with a high efficiency HPGe detector, for the measurement of the activity resulting from irradiation of a sample in the NEAR station. The use of a moderator/filter assembly is envisaged, in order to produce a neutron beam of Maxwellian shape at different thermal energies, necessary for the measurement of Maxwellian Averaged Cross Sections of astrophysical interest. A new fast-cycling activation technique is also being investigated, for measurements of reactions leading to isotopes of very short half life.

# 1 Introduction

Neutron-induced reactions play an important role in a variety of fields in fundamental and applied Nuclear Physics. In nuclear astrophysics, neutron capture reactions and $\beta$-decay are responsible for the synthesis of all elements from iron up to uranium in the Universe, in two main stellar scenarios, each being responsible for the production of about one half of the observed elemental abundances. The slow neutron capture process (or *s* process in short), occurs during the advanced burning phases of stellar evolution [1]. Depending on the stellar mass, the *s* process operates in thermally pulsing low-mass Asymptotic Giant Branch (AGB) stars (main component) [2] or during core He and shell C burning in massive stars (weak component) [3]. Rapid neutron capture, or *r* process [1], occurs in explosive scenarios and leads to the production of short-lived and very neutron-rich nuclei, which later decay to the known stable isotopes. Recent works suggest that a likely source of *r*-process elements could be the catastrophic aftermath of neutron star mergers [4].

The main features of the chemical evolution of the Universe and of heavy element production, are reasonably well described by stellar models incorporating the two neutron capture processes mentioned above. To this end, accurate stellar cross sections are a fundamental input in order to reproduce the observed abundances. At present, the knowledge of neutron cross sections is still unsatisfactory for several isotopes.

Neutron-induced reactions play a key role also in a variety of applications, related to the fields of energy, nuclear medicine, material studies, as well as reactor, environmental and space dosimetry. In particular, data on neutron capture and neutron-induced fission reactions are important for improving safety operation of current fission reactors or for the design of future generation ones, including systems for nuclear waste incineration [5]. In nuclear medicine, neutron-induced reactions are at the basis of the production of a number of radioisotopes for diagnosis and therapy, and the knowledge of the cross section for a series of neutron reactions in biological tissues is fundamental for an accurate dose estimate in radiotherapy.

Neutron cross section data are also needed for the lifetime assessment of fusion reactors now being developed [6]. As an example, integral cross sections are required for a number of elements (or specific isotopes) for the estimate of the radiation damage in structural material of the reactors. In particular, integral cross sections for both transmutation and (n,cp) reactions, are needed to estimate neutron damage effects in the inner components of the reactors, related to the variation of the chemical composition of the material and/or to gas production [7]. Activation cross sections, mostly related to neutron capture reactions, are also needed to address safety, licensing, decommissioning, and waste management issues [8]. At present, cross section data in this respect are either missing or unreliable for several isotopes, both below and above 14 MeV, the energy of neutrons produced in D-T generators. The request of data at high energy is related to material studies at future neutron irradiation facilities, such as IFMIF and its precursor DONES [9], characterized by neutron spectra that extend up to several tens of MeV. As in the case of astrophysics-related requests, the lack of data is a direct consequence of the lack of suitable neutron facilities, in terms of energy and flux of the beam, as well as of the typically low cross section or unavailability of isotopically pure samples.

Energy-dependent cross section of neutron-induced reactions relevant to nuclear astrophysics and applications have been measured for more than two decades at the n_TOF facility at CERN [10] by the time-of-flight technique. However, the intrinsic limitations of the technique, in terms of signal-to-background ratio, has hindered much-needed measurements for a number of important isotopes, in particular short-lived unstable ones, and for specific reaction channels.

To overcome such limitations, a new experimental area, the NEAR station, set at a very short distance from the spallation source, has recently been proposed and constructed at n_TOF during the long CERN accelerator shut down (2019-2021). A detailed description of this newly built area is given in [11]. The purpose of this work is to present the main features of the neutron beam in the NEAR station and its capabilities with respect to the implementation of the activation technique. The new experimental area is complemented by a dedicated laboratory equipped with HPGe detectors, referred to as the GEAR station (Gamma spectroscopy Experimental ARea), for $\gamma$-ray and $\beta$-particle spectroscopy.

The new high-flux irradiation station for activation measurements (referred in the following as the a-NEAR station) will allow to fully exploit the sensitivity and the selectivity of the activation technique, making feasible measurements of neutron activation cross section for reactions that until now have been outside the possibilities of n_TOF, finally allowing to experimentally access previously unexplored physics cases.

Specifically, by means of a neutron moderator/filter system, Maxwellian-like neutron spectra could be produced to be used for measurements of Maxwellian Averaged Cross Sections (MACS) at different stellar temperatures, on unstable and low mass samples or for cases in which extremely low reaction rates are expected. Additionally, concerning future plans, the CERN-PS proton pulse time structure (one pulse every 1.2 s) may allow the implementation of the fast cyclic activation technique [12]. In this paper we present the main features of the newly completed a-NEAR station, the characteristics of the neutron beam, the laboratory for activity measurement, and new measurement capabilities currently being investigated.

The paper is organized as follows: in Section 2 the a-NEAR station is described. The main features of the neutron beam in terms of neutron spectrum, flux and spatial profile, provided by Monte Carlo simulations are presented in Section 3. The $\gamma$-ray spectroscopy laboratory for activity measurement is discussed in Section 4. Section 5 presents two techniques now being investigated for astrophysics-related measurements. Conclusions and perspectives are finally given in Section 6.

[a] nicola.colonna@ba.infn.it

## 2 The NEAR experimental area

During CERN's Long Shutdown period of 2019-2021, the n_TOF facility has undergone major updates. The target pit shielding has been completely overhauled and a new generation spallation target has been installed [13]. Taking advantage of the consolidation work on the n_TOF target area, a new experimental station, generically referred to as the 'NEAR' Station (because of its vicinity to the spallation target), has been constructed. The main purpose of this new installation is to exploit the extremely high neutron intensity close to the target in order to perform challenging activation measurements as well as to study radiation damage on non-metallic materials [14]. The NEAR Station comprises two areas, the NEAR Irradiation Area (indicated as i-NEAR) and the NEAR Activation Area (a-NEAR), with the first positioned directly next to the spallation target and the second being located outside the target pit shielding, at a distance of 3 m from the target center. The details of i-NEAR can be found in [11], while this paper focuses on the features and applications of the a-NEAR. A schematic design of the a-NEAR station, as well as a side view of the spallation target and shielding area can be seen in Fig. 1.

present (brown wall in Fig 1). The room has a rectangular shape, 7 m long, 2.6 m wide and has a height of 2.8 m. The beam is transported through a 26 cm diameter hole in the shielding wall, at 1.2 m from the floor, specifically designed to house a collimation system. The collimator is 80 cm long and is placed just inside the middle section of the shielding wall, made of concrete. It has a conical shape, with an entrance and exit holes of 25 and 6 cm diameter, respectively. The first 50 cm are made of stainless steel, while the remaining 30 cm are made of a series of borated polyethylene disks. A photo of the collimator exit face is shown in Fig. 2, along with a sample holder used for activation measurements. Neutron irradiation is performed just outside the marble wall, at 20 cm distance from the exit hole of the collimator, inside the a-NEAR room. The neutron beam is dumped on the wall opposite the collimator exit, where irradiation of additional material (such as electronic devices) can be performed. The neutron beam collimation system points to the side of the target (orthogonal to the proton beam direction). As a consequence, the very high energy component of the neutron beam is absent. An additional moderation of the neutron beam can be performed by placing a block of suitable moderating material close to the target. In this respect, a shelf for housing such a moderator has been designed and implemented on the structure currently used for material irradiation studies (see Figure 6 in Ref. [11]). The importance of the moderator, together with a low-energy filter, is discussed in Section 5.1.

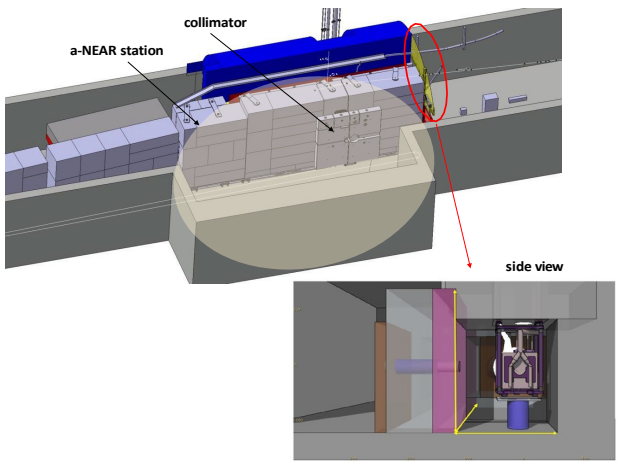

**Fig. 1.** The NEAR Station schematic design (top), and side view of the spallation target area, as implemented in simulations of the n_TOF facility.

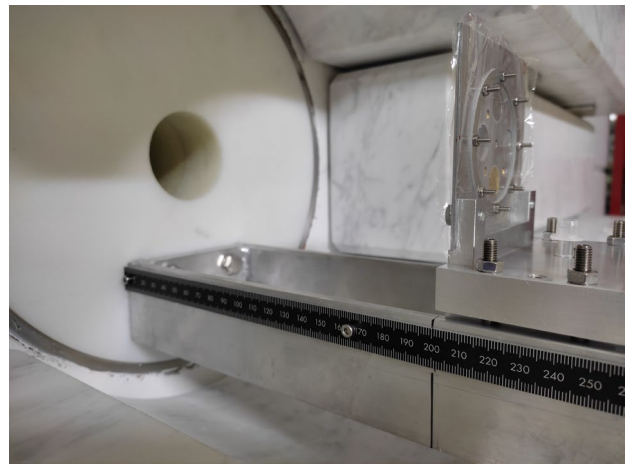

**Fig. 2.** The neutron irradiation position as seen from inside the a-NEAR station. The photo shows the collimator exit surface, the last part of the shielding wall, made of marble, and a multi-foil sample holder used for the commissioning of the a-NEAR neutron beam, mounted on a rail.

The a-NEAR experimental area is located just outside the shielding wall on the left side of the n_TOF neutron spallation target (as seen from the proton beam). The shielding is made of 40 cm thick iron wall followed by 80 cm thick concrete (shown as purple and gray walls in the bottom panel of Fig. 1). In correspondence of the neutron beam exit, an additional 20 cm thick marble shielding is

One of the advantages of the a-NEAR Station is that irradiation with the neutron beam can be performed parasitically, i.e. without disturbing experiments simultaneously performed in the other two experimental areas of

n_TOF (EAR1 and EAR2). The main advantage of the new area, however, is related to the very high neutron flux, direct consequence of the short distance from the spallation target. In the following, the main features of the neutron beam, determined by means of Monte Carlo simulations, are presented.

## 3 The a-NEAR neutron beam

fluka Monte Carlo simulations [15–17] were carried out to characterize the area around the collimator, in terms of particle fluence and dose level. The neutron 2D spatial distribution was recorded in two positions: right at the exit of the collimator (hereafter denoted as position "A"), and at a distance of 20 cm from the exit face of the collimator, aligned with the external surface of the marble shielding (denoted as position "B").

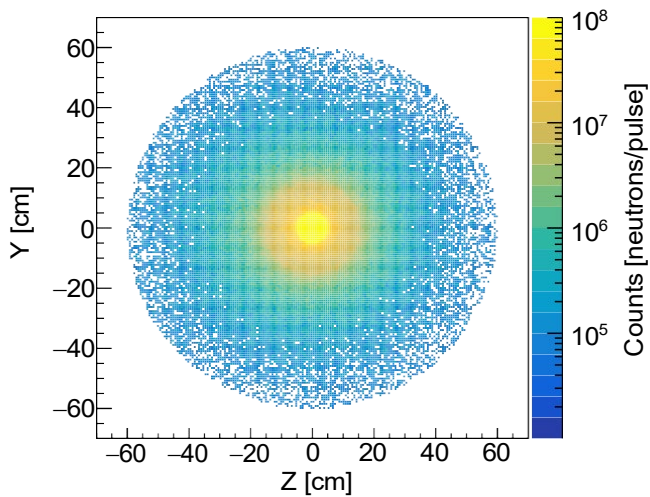

**Fig. 3.** 2D neutron distribution on the external marble surface (position "B"), with binning of 1x1 $cm^2$ and within 60 cm of radius from the center of the collimator.

According to simulations, this last position represents the best compromise between beam dimension and neutron background, and therefore has been chosen for the irradiation of most of the samples. In the simulations, neutrons are transported from the side face of the spallation target through the collimator and the shielding wall, and recorded on a scoring surface of 60 cm radius. The resulting beam spatial profile is shown in Figure 3. The plot shows the neutrons crossing the scoring surface from all directions, normalized to the nominal proton pulse (one pulse corresponding to $7 \times 10^{12}$ protons). Secondary hadrons other than neutrons (i.e. protons, muons, etc.) exiting the collimator amount to less than 0.3% of the total number of particle crossing the scoring surface and can be neglected.

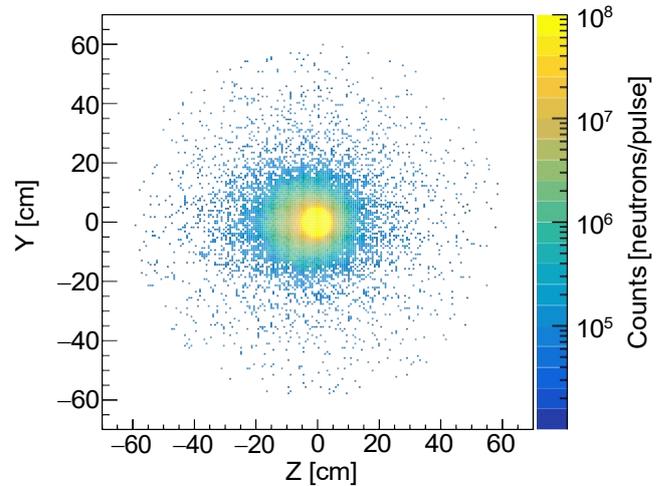

**Fig. 4.** 2D neutron distribution on the external marble surface (position "B"), with binning of 1×1 $cm^2$ and within 60 cm of radius from the center of the collimator, crossing the scoring surface at an angle within 10 degrees relative to the normal.

Figure 4 shows the neutron distribution on the same surface as for Figure 3, but considering only neutrons impinging within an angle of 10 degrees with respect to the normal to the scoring surface. As can be seen, the distribution is not symmetric anymore with respect to the collimator center and presents a higher flux for negative values of the Z-axis, corresponding to the left of the collimator as seen from the target. The shift is due to the geometry of the collimator with respect to the target, which is not perfectly perpendicular but exhibit an angle of 5 degrees. This effect is not visible in Figure 3 probably because of the neutron scattering around the collimator and shielding, when no condition of the crossing angle is applied.

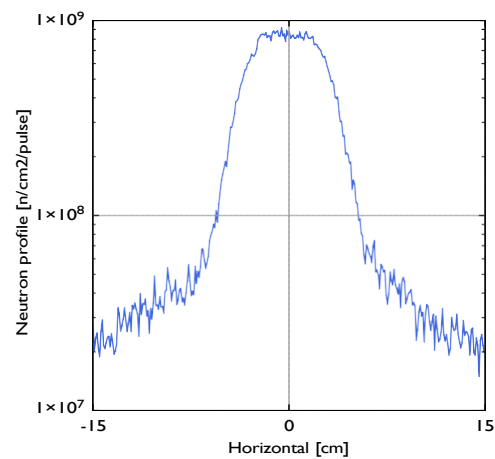

**Fig. 5.** Neutron beam profile in horizontal direction on the external marble surface (position "B").

Figure 5 shows the projection of the neutron beam profile on the horizontal direction (Z) along the exit face of the collimator (that has a radius of 15 cm). From the figure it can be appreciated that the neutron flux is as high as $9\times10^8$ n/cm$^2$/pulse. The figure also shows that outside the collimator hole, a tail of neutrons is present, mostly produced by neutrons scattered inside the collimator itself or in the shielding wall.

Figure 6 shows the simulated neutron spectra for different conditions on the spatial profile reported in Fig. 3. Specifically, the condition is applied on the radius $R_B$, from the collimator center, on the marble surface (position "B"). For instance, the spectrum denoted as $R_B < 15$ cm includes only neutrons recorded on the scoring surface within 15 cm distance from the collimator center. It should be noticed that the spectra are shown in units of counts to allow a direct comparison, hence, they are not normalized per unit surface. For comparison, the spectrum on the internal marble surface (position "A") is also shown in the figure, with $R_A$=15 cm, the radius of the collimator. The comparison indicates that the neutron flux at that position is slightly larger than the one on the surface B, for the same radius, as expected from beam divergence considerations.

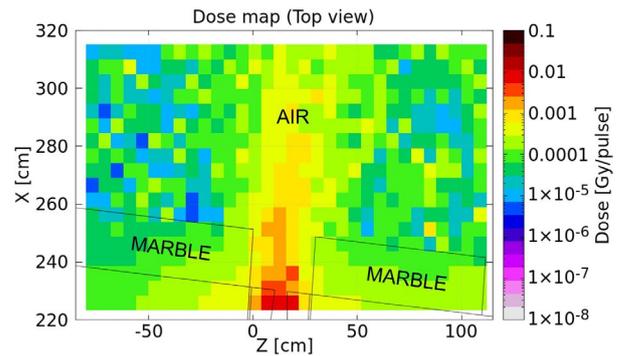

**Fig. 7.** Dose map overview at the exit of the collimator. The target is located towards lower X values.

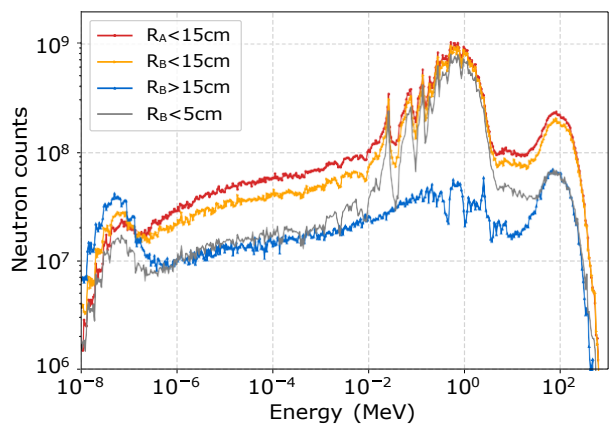

**Fig. 6.** Neutron spectra at the exit of the collimator (whose maximum radius is 15 cm) scored on the surface just at the exit of the collimator hole, (position A) and at the external surface of the marble shielding (position B), considering different radii.

Another consideration regards the spectrum of neutrons recorded in the position B "outside" the collimator, i.e. at a distance from the collimator center greater than 15 cm ($R_B > 15$ cm). Such neutrons are at least partially traversing the marble shielding and represent a source of background outside the neutron beam. It can clearly be seen in Figure 6 that neutrons of energy above 0.2 eV are strongly attenuated, while on the contrary the thermal neutron flux is larger because of the thermalization of fast neutrons when passing through the shielding.

The total ionising dose level recorded outside the collimator, averaged in a 30 cm wide horizontal plane centered on the collimator is shown in Figure 7. The meaningful information in this plot is the dose in air, as the marble is part of the shielding wall and it is therefore inaccessible. The dose values in air in the proximity of the collimator exit face are around 1-2 mGy/pulse. This relatively large value implies that it may be difficult to operate detectors (either active or passive ones) in the vicinity of the collimator, unless they are characterized by a high radiation tolerance or are very carefully shielded. The figure also shows that, like for neutrons, photons as well exit the collimator slightly shifted on one side with respect to the center.

## 4 The GEAR station

Activation measurements at n_TOF are carried out in two steps. In the first one, the sample is irradiated with the neutron beam in the NEAR station, right outside the shielding wall in correspondence of the hole of the collimator (see Figure 2). The second step of the activation technique consists in the accurate measurement of the induced sample activity. To this purpose, the n_TOF Collaboration has setup a dedicated experimental area, the GEAR Station. This is located at the premises of the n_TOF facility at CERN in an underground area. The low radiation background conditions as well as the fact that this is a radiation controlled area allows for the safe handling of irradiated samples.

The GEAR station is equipped with a CANBERRA HPGe detector GR5522. Photos of the setup are shown in Figures 8 and 9. The detector is a n-type HPGe detector of 63% relative efficiency. This feature, combined with a very thin carbon composite window (or end cap), allows for high efficiency measurements of γ-rays with energy from 3 keV up to the 10 MeV, considerably pushing down the low energy limit. The energy resolution of the detector is 3 keV at 1.33 MeV γ-ray energy, with a peak-Compton ratio of 64:1. The preamplifier circuit of this detector is equipped with a suitably designed fast-switch circuit [18] that allows to use the detector also for time-of-flight measurements close to the neutron beam line in the other two n_TOF experimental areas (EAR1 and EAR2) where a large prompt signal, or γ-flash, is present. For the same reason, the detector is cooled by means of the CANBERRA CP-5 electrically refrigerated cryostat that allows for continuous and safe operation of the HPGe detector in cases where large volume LN2 Dewars have to be avoided (in particular in underground areas with limited space). The performance of an electrically refrigerated device is similar to the one of a liquid-nitrogen cooled HPGe.

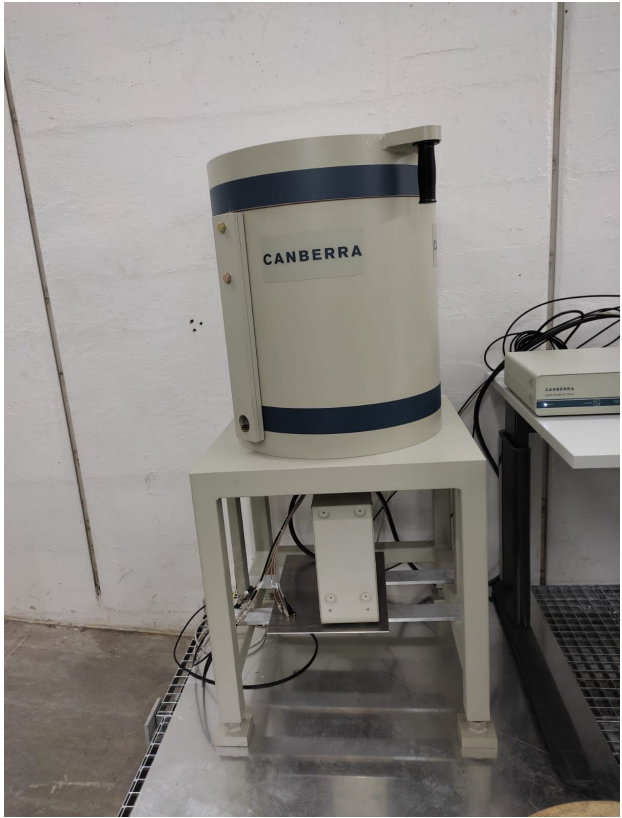

**Fig. 8.** The HPGe-based γ-ray spectrometer, placed in the GEAR Station. The detector is inside the lead shielding shown in the photo, while the refrigeration and front-end electronics are located below the shielding.

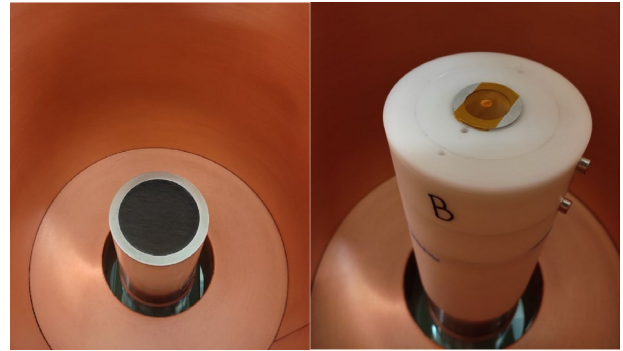

**Fig. 9.** Left panel: A view from the top of the HPGe detector inside the lead shield. Right panel: same as before, but with the polyethilene sample holder on top of the detector.

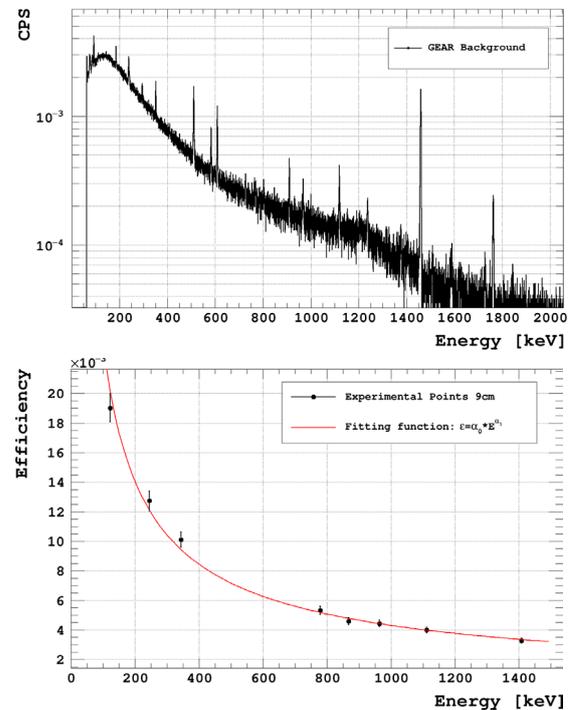

**Fig. 10.** Top panel: The background of the HPGe detector in the GEAR station. Bottom panel: the efficiency of the γ-ray spectrometer measured with calibrated sources for a detector-source distance of 9 cm. The red curve represents a fit of the measured efficiencies with a power law, used for interpolation purposes. See text for more details.

For optimal background conditions, the HPGe device is placed inside the 747 CANBERRA lead shield, shown in Figure 8. The HPGe detector inside the lead shield is shown in the left panel of Figure 9.

Measurements of the activity of the irradiated samples can be performed at different source-to-detector distances, ranging from 0 - 25 cm, thanks to a suitably designed system made of different polyethilene spacers and sample holder that can be stacked one on top of the other. A picture of such a system, mounted on the detector, is shown in the right panel of Figure 9. The versatile detection geometry is instrumental for high accuracy measurements and can be chosen depending on the observed (or expected) counting rate. The reproducibility of the detection geometry is a significant factor that defines the efficiency of the detector and accordingly the accuracy of the experimental results. Thanks to the spacers system the distance between the sample and the detector window can be reproduced with an accuracy better than 0.1 mm.

The most important requisites for high accuracy measurements of the activity are a low background and the reliable knowledge of the detector efficiency as a function of the γ-ray energy. This goal is achieved by means of several calibration γ-ray sources, whose activity is known to within 1-2 % accuracy. Figure 10, top panel, shows the background measured for a threshold on the amplitude corresponding to a deposited γ-ray energy of 63 keV. The bottom panel shows the efficiency measured with a $^{152}$Eu calibrated point-like source, with γ-rays of energy spanning from 100 keV to 1.4 MeV, at a distance from the de-tector of 9 cm. For interpolation purposes, the measured efficiency was fitted with a power law shown by the red curve in Figure 10.

Beside the standard sample holder, a special setup is foreseen that will be equipped with a plastic cylindrical BC408 detector of 4 cm in diameter and 10 mm thickness. In this way the sample will be "sandwiched" between the HPGe and the plastic detectors in close geometry. For $\beta$-only emitters the thin carbon composite window of the HPGe detector allows the detection of $\beta$s when the minimal source-to-detector distance is adopted, while the efficiency can be further extended and controlled through the plastic detector. Furthermore, with the same configuration $\beta$-$\gamma$ coincidence measurements can be utilized that could improve further the signal to background ratio with respect to the inclusive $\gamma$-ray measurement.

For the GEAR station two Data Acquisition systems are available. When standard singles measurements are performed a system based on 2026 Mirion Amplifier and a Digital Multichannel Analyzer - Amptek MCA-8000D is used. Additionally, a fully digital DAQ system is available based on the CAEN DT 5730S 500MS/s 8-channel digitizer. The digital DAQ extends significantly the flexibility of data analysis modes (singles or coincidences) by recording the full waveform of the signals from the HPGe preamplifier and the anode signal of the plastic scintillator photomultiplier tube.

## 5 New measurement capabilities and perspectives

In this section we present two examples of activation measurements that can be performed at the a-NEAR station. Both cases are mostly relevant for an experimental program on nuclear astrophysics, but the use of beam-shaping techniques and innovative instrumentation could make feasible challenging measurements of interest for other fields as well, in particular for activation measurements relevant for fusion technologies or for nuclear medicine.

### 5.1 Beam filtering for astrophysics-related measurements

One of the most important nuclear data needs for the understanding and modelling of stellar nucleosynthesis is the Maxwellian Averaged Cross Section of neutron capture reactions, for various thermal energy (kT) values, typically ranging from a few keV to a few hundred keV. Two methods are typically used to experimentally determine this quantity: i) measure the energy-dependent reaction cross section and subsequently calculate the MACS by performing a spectral averaging using a Maxwellian spectrum of the desired temperature; 2) measure the integral cross section by irradiation with a neutron beam with Maxwellian-like spectrum. While this last method has been, and it is still being, used at neutron facilities based on low-energy accelerators [19–21], at n_TOF all MACS of astrophysical interest have so far been obtained from time-of-flight measurements. As already mentioned, however, the limited sensitivity of this technique has up to now hindered the possibility to investigate some challenging reactions, in particular on short-lived radionuclides. Whenever applicable, the highly-sensitive activation technique can enable access to isotopes available in very small quantities, as small as a few ng, that are difficult to access in a TOF experiment. The construction of the NEAR station has opened the possibility to directly measure at n_TOF the MACS of various isotopes, at different thermal energies, by means of the activation technique. To this end, the neutron spectrum has to be suitably shaped so to resemble a Maxwellian-like distribution. This can be achieved by means of a moderation system, needed in particular to suppress the high-energy component, to be placed right next to the spallation target, and a filtering setup to suppress the low-energy component, that can be installed in the a-NEAR station, just downstream of the collimator. Different choices of the moderation material are now being investigated, for different applications. An accurate reproduction of a Maxwellian spectrum and a high flux can be obtained with a moderator of suitable material and dimension. Analytical calculations, confirmed by detailed Monte Carlo simulations performed with the geant4 toolkit (described below), show that indeed a reasonable shaping of the neutron spectrum can be achieved by using an AlF$_3$ moderator and a $^{10}$B-based filter. An interesting feature at the NEAR station is that Maxwellian-like neutron spectra corresponding to different thermal energies $kT$ can be obtained simply by varying the thickness of the $^{10}$B-based filter. The movie in Figure 11 shows the effect of filters of different thicknesses, combined with an AlF$_3$ moder-ator of 40 kg mass. In each frame, a comparison of the a-NEAR spectrum with a Maxwell-Boltzman distribution at the thermal energy that best fits the a-NEAR spectrum is shown. The similarity between the two spectra is evident in particular for the stellar temperatures around 25 keV. Preliminary calculations indicate that MACS at different temperatures can be determined by the activation technique in the NEAR station, up to about kT=100 keV, with reasonable accuracy. Figure 12 provides some indication on the uncertainty related to the use of the technique described above for MACS determination. For each isotope, the figure shows the ratio between the Spectral Averaged Cross Sections (SACS) calculated either by convoluting the currently known evaluated cross section with the maxwellian-like neutron spectrum in the a-NEAR station, or with the true Maxwellian distribution, at kT = 25 keV. It can be seen that the ratio is in average close to 1, with differences of the order of 10-15%, except for the lightest isotopes. This value can therefore be assumed as a realistic estimate of the most probable uncertainty in the MACS that would be determined in a measurement at the a-NEAR station for any isotope of unknown capture cross section.

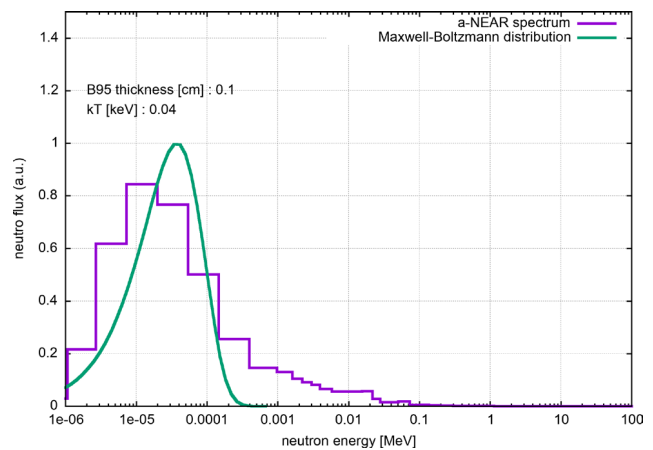

**Fig. 11.** Animation: The neutron spectrum in the a-NEAR station, obtained with the use of an AlF$_3$ moderator of 40 kg mass, placed next to the spallation target and a $^{10}$B-based filtering system placed at the exit of the collimator (purple histogram). The different frames of the animation show the spectrum for increasing filter thickness. For comparison, the green curve shows the Maxwell-Boltzmann distribution at a given thermal energy kT (shown in the legend), that best fits the a-NEAR spectrum.

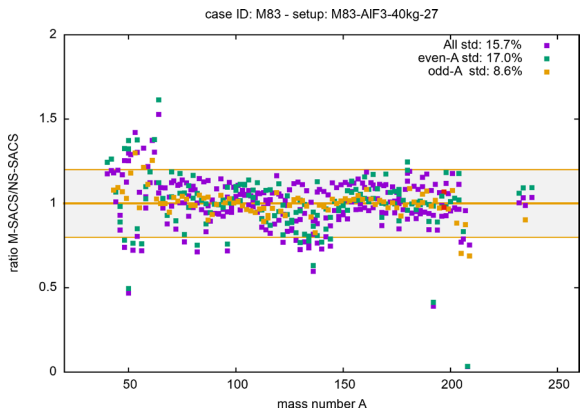

**Fig. 12.** Ratio between the Spectral Averaged Cross Section (SACS) calculated by convoluting the capture cross section from ENDF/B-VIII of each isotope with the maxwellian-like neutron spectrum in the a-NEAR (40 kg $AlF_3$ moderator plus 3 cm thick boron filter 95% $^{10}$B-enriched) and the SACS calculated with a corresponding Maxwell-Boltzmann distribution (kT=25 keV).

The use of a moderator to suppress or reduce the high-energy component may not be strictly necessary for astrophysics-related measurement, since high-energy neutrons ($E_n$ >1 MeV) do not contribute significantly to the MACS, due to the typically negligible value of the capture cross section at high neutron energies. A filtering system alone could therefore be enough for measurements of the MACS, at least in a restricted thermal energy range. Such an option is now being investigated, both with simulations and experimentally. The neutron energy distribution at the irradiation position can be estimated by means of Monte Carlo simulations, performed with the geant4 toolkit, assuming a sample placed in the center of a cylinder made of boron carbide ($B_4C$) highly enriched in $^{10}$B. Neutrons in the thermal and epithermal energy regions are moderated and eventually absorbed by the $B_4C$ cylinder, producing a peak in the keV region that resembles a Maxwellian distribution, plus a tail at high energy that, however, does not contribute significantly to the SACS. While some information on the MACS can be obtained in this way, an upgrade of the NEAR station is already foreseen, with the installation of a moderator inside the spallation target pit, in order to improve the agreement between the filtered and expected Maxwell-Boltzmann distributions.

### 5.2 The CYCLING station

The high neutron flux at a-NEAR opens the way to challenging (n,γ) activation measurements on extremely small mass samples and on radioactive isotopes. Thus far, samples are irradiated for several days to weeks at a-NEAR and transported to and measured in the GEAR station described in Sec. 4. This activation scheme is well suited for (n,γ) measurements with long-lived unstable product isotopes (i.e. $t_{1/2} \geq$ several hours). However, it imposes an intrinsic limit on the accessible physics cases, specifically on the minimum half-life of the (n,γ) product, given by the required time to cool down the a-NEAR bunker and to transport the sample to the offline detector. This limitation can be overcome with the installation of a future cyclic activation station for (n,γ) experiments (CYCLING). This setup would enable the repetition of a short irradiation, rapid transport to a detector, measurement of the decay and transport back to the irradiation position. Depending on the duration of the transport of the sample to the detector, half-lives of the order of seconds become accessible. This process is repeated for a number of cycles, and with the cyclic activation analysis technique [22] the spectra from each counting cycle are summed together to give one final total spectrum. By means of this approach, the counting statistics and signal-to-background ratio for short-lived nuclei of interest can be considerably increased.

There are several isotopes of great interest for stellar nucleosynthesis studies that will be investigated at GEAR and the new CYCLING station. Some of them are stable nuclei, already known in certain neutron-energy ranges, and thus they will be measured at the beginning of the experimental campaign for benchmarking the performance of the CYCLING station in terms of accuracy and sensitivity. One such case is $^{19}$F, whose neutron capture cross section is of great interest for understanding the origin of fluorine in the solar system [23]. Neutron capture on $^{19}$F leads to the formation of $^{20}$F with a half-life of only 11 s. Given the very small neutron capture cross section of only few mb in the keV energy range, and the short half-life of the product nucleus, this type of nuclei are excellent candidates for the cyclic activation technique. $^{19}$F is copiously produced in AGB stars and an excess of $^{19}$F has been observed in this type of stars [24], which actually led to the measurement of this cross section at $kT$ = 25 keV at FZK using also the cyclic-activation technique [25]. The MACS measured at 25 keV has an uncertainty of less than 3%, thus representing a good case for validating the performance of the CYCLING station. Once validated, it will become possible at CYCLING to extend the measurement of the 19F MACS to other energy ranges of interest in AGB stars, for example at $kT \sim 8$ keV, which represents also a significant contribution to the destruction of 19F in AGB stars [23].

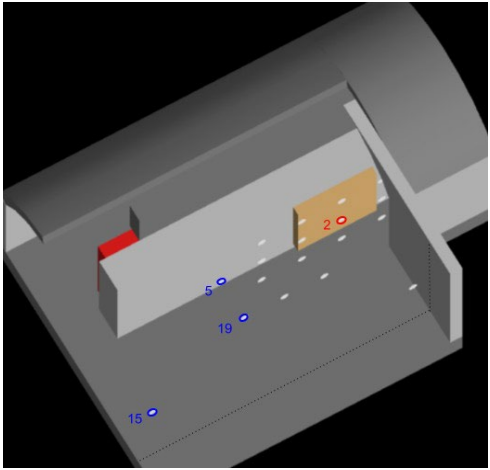

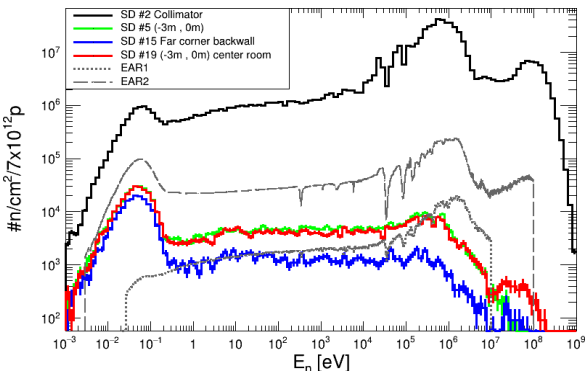

**Fig. 13.** Top: view of the a-NEAR bunker as implemented in geant4 (top). Bottom: Neutron fluence of the n_TOF beam lines (EAR1/2 and a-NEAR) compared to the expected off-beam neutron fluence registered in the highlighted scorers far from the a-NEAR beam.

From the astrophysics viewpoint, perhaps one of the most striking applications of the CYCLING station will be the possibility to directly access, for the first time, stellar $(n, \gamma)$ measurements of relevance for the intermediate (*i*) neutron capture process [26]. The *i* process involves neutron capture at neutron densities of $10^{13}$–$10^{16}$ cm$^{-3}$, in between the *s* and *r* processes. Recently, the *i* process attracted significant interest because it might explain the abundance pattern of a special kind of Carbon-Enhanced Metal-Poor stars (CEMPs), called CEMP-s/r [27]. From an experimental standpoint, the requirement of target-nuclei with short half-lives, such as some of those involved in the *i* process, leading after neutron capture to product nuclei with even shorter half-lives, of the order of 1-to-100 s, offers a unique opportunity for combining in a synergetic fashion CYCLING with the nearby ISOLDE facility.

For example, it has been found that in the *i*-process conditions variations in the neutron-capture rates of some specific isotopes could affect the observable elemental ratio predictions involving Ba, La and Eu in *i*-process conditions [28]. In this respect, some of the most relevant neutron-capture studies for future experiments at CYCLING could be [29]:

137Cs(30 a)$(n, \gamma)$138Cs(32.2 m) and
144Ce(285 d)$(n, \gamma)$145Ce(3 m).

The feasibility of CYCLING will depend on various factors. Firstly, the time between consecutive proton bunches ($\geq 0.8$ Hz), would provide periods between consecutive pulses of up to various seconds. This condition is well suited for the measurement of the decay of very short-lived isotopes. Moreover, a high-resolution $\gamma$-ray detector (e.g. HPGe) should be operated in the harsh radiation environment of the a-NEAR and thus the design of the measuring station poses a substantial challenge due its close proximity to the spallation target.

In this context, an excellent knowledge of the expected neutron and $\gamma$-ray fields is a prerequisite for its feasibility. The first step of this study has been carried out by means of Monte Carlo simulations of the expected off-beam neutron flux (or more precisely, the neutron intensity) at various positions of the a-NEAR bunker (see top panel of Fig. 13). This simulations have been carried out with the geant4 toolkit (v10.6 p.03) using as primary particles the neutrons scored in the fluka simulations of Sec. 4. The average neutron flux per proton pulse at the collimator exit in a-NEAR is compared with three off-beam positions providing the lowest flux (blue scorer in the Fig. 13). The comparison of the latter with the flux of the neutron beam in the first (EAR1) [30] and second (EAR2) [31] experimental areas at n_TOF indicates that the off-beam neutron flux at a-NEAR is in between the ones in EAR1 and that in EAR2. In contrast to the direct beam in a-NEAR, the off-beam spectra are dominated by back-scattered neutrons, typically below the 10 MeV range and feature an enhanced thermal component.

Following this simulation study, an experimental characterization of the neutron and $\gamma$-ray field in positions of interest for the design of the CYCLING system needs to be carried out, with different active and passive detectors, and it is already planned [32]. The conclusions of these measurements will shed light on the background conditions of the a-NEAR and help in the optimization of the dump design and shielding elements that will facilitate the installation of the future CYCLING system.

# 6 Conclusions

The n_TOF experimental capability has recently been increased with the construction of a new high-flux experimental area for activation measurements. The new installation, called a-NEAR, exploits the extremely large neu-tron flux at 3 m from the spallation target. The neutron beam enters the new station though a hole in the target shielding, in which a collimator system in inserted. Given the short distance from the target, the new station will be mostly dedicated to integral cross section measurements, either with the original neutron beam or with a moderated/filtered beam, specifically designed to suitably shape the neutron spectrum. In particular, Monte Carlo simulations indicate that the use of neutron filters of different thickness allows one to obtain maxwellian-like spectra of different thermal energy, to be used for challenging measurements of the MACS of short-lived isotopes, of interest for nuclear astrophysics, with a reasonable uncertainty. The large flux and wide energy range of the neutron beam makes also feasible activation measurements of interest for energy and medical applications. The a-NEAR station is complemented with a high-efficiency, low-background $\gamma$-ray and $\beta$-particle spectroscopy system, based on HPGe detectors, positioned inside a dedicated room nearby the irradiation station. Future possibilities are now being investigated to perform fast-cycle irradiation-activation measurements inside the a-NEAR, using an off-beam HPGe-based detection system. A first campaign for the commissioning of the neutron beam in the a-NEAR station has been performed, confirming the predictions of Monte Carlo simulations in terms of neutron flux, beam profile and background. The results of the commissioning will be reported in a forthcoming paper.